\documentclass[12pt]{article}
\usepackage{amssymb,amsmath,epsfig}

\begin{document}

\title{\bf Gravitational Perfect Fluid Collapse in $f(R)$ Gravity}

\author{M. Sharif \thanks{msharif@math.pu.edu.pk} and H. Rizwana
Kausar
\thanks{rizwa\_math@yahoo.com}\\
Department of Mathematics, University of the Punjab,\\
Quaid-e-Azam Campus, Lahore-54590, Pakistan.}

\date{}
\maketitle

\begin{abstract}
In this paper, we investigate spherically symmetric perfect fluid
gravitational collapse in metric $f(R)$ gravity. We take non-static
spherically symmetric metric in the interior region and static
spherically symmetric metric in the exterior region of a star. The
junction conditions between interior and exterior spacetimes are
derived. The field equations in $f(R)$ theory are solved using the
assumption of constant Ricci scalar. Inserting their solution into
junction conditions, the gravitational mass is found. Further, the
apparent horizons and their time of formation is discussed. We
conclude that the constant scalar curvature term $f(R_0)$ acts as a
source of repulsive force and thus slows down the collapse of
matter. The comparison with the corresponding results available in
general relativity indicates that $f(R_0)$ plays the role of the
cosmological constant.
\end{abstract}


\section{Introduction}

New evidence coming from astrophysics and cosmology has revealed a
quite unexpected picture of the expanding universe. This cosmic
acceleration is directly supported by high red-shift supernovae,
large scale structure and weak lensing. What causing the repulsive
effect in the cosmic expansion is one of the mysteries of modern
era. Theoretically, the phenomenon can be accounted for either by
supplementing the energy-momentum tensor by exotic matter component
with large negative pressure (dark energy) \cite{1}-\cite{3} or by
modifying gravity itself. There exist a family of modified theories
of gravity which extend in some way general relativity, for example,
scalar tensor theory of gravity, Brans-Dick theory, string theory,
Guass-Bonnet theory and $f(R)$ theories of gravity.

The $f(R)$ theories of gravity provide the very natural
gravitational alternative for dark energy. It has given birth to a
great number of papers in recent years, for example,
\cite{1*}-\cite{3*}. Several features including solar system test
\cite{4*}, Newtonian limit \cite{5*}, gravitational stability
\cite{6*} and singularity problems \cite{7*} are exhaustively
discussed in this theory. General principles such as the so-called
energy conditions have been used to place constraints on $f(R)$
theory \cite{7**}. Amendola \textit{et al}. \cite{6} have derived
the conditions under which dark energy $f(R)$ models are
cosmologically viable. Erickcek \textit{et al}. \cite{7} found the
unique exterior solution for a stellar object by matching it with
interior solution in the presence of matter sources in metric $f(R)$
gravity. Kainulainen \textit{et al.} \cite{8} studied the interior
spacetime of stars in Palatini $f(R)$ gravity.

Gravitational collapse is one of the most important and thorny
problems in general relativity. It is at the heart of structure
formation in the universe through which clusters of galaxies,
stellar groups, stars and planets came into being. One can study the
gravitational collapse by taking the interior and exterior regions
of spacetime. The proper junction conditions are formed that allow
the smooth matching of these regions. Oppenheimer and Synder
\cite{os} were the pioneers to study gravitational collapse. Herrera
and Santos \cite{Herr} discussed the dynamics of gravitational
collapse which undergoes dissipation in the form of heat flow and
radiation.

It has been predicted that gravitational collapse of massive objects
leads to the formation of spacetime singularities
\cite{10**},\cite{11*}. These singularities can be of two types: an
observable singularity (naked singularity) and a singularity that
cannot be observed (black hole). Penrose \cite{11} suggested a
cosmic censorship conjecture (CCC) according to which the final
outcome of a gravitational collapse must be a black hole. Most of
the studies in gravitational collapse have considered the
Tolman-Bondi-Lemaitor spherically symmetric spacetimes containing
irrotational dust to support or disprove the CCC
\cite{CC},\cite{CCC}. A general conclusion from these studies is
that a central curvature singularity forms but its local or global
visibility depends on the initial data.

An extensive study of gravitational collapse has been carried out
in the presence of cosmological constant. Markovic and Shapiro
\cite{MS} generalized the pioneer work in the presence of a
positive cosmological constant. Lake \cite{Lake} extended this
study with both positive and negative cosmological constants.
Sharif and Ahmad \cite{12} investigated gravitational collapse of
a perfect fluid with a positive cosmological constant and
generalized this work to five and higher dimensional spacetimes
\cite{hd}. Recently, Sharif and Abbas \cite{sa} extended the same
work for electromagnetic charged perfect fluid. In this paper, we
explore gravitational collapse of a perfect fluid in metric $f(R)$
gravity.

The paper is organized as follows. Junction conditions are discussed
in the next section. In section \textbf{3}, we solve the field
equations in metric $f(R)$ gravity. Here we assume the condition of
constant scalar curvature which is very large to meet the
requirement of gravitational collapse. Section \textbf{4} is devoted
to study the apparent horizons and their time of formation. We
summarize and discuss the results in section \textbf{5}.

\section{Junction Conditions}

In this section, we derive the conditions to be satisfied at the
surface of a collapsing perfect fluid sphere. We assume spherical
symmetry about an origin $O$ and a $3D$ hypersurface $\Sigma$
centered at $O$ which divides spherically symmetric spacetime into
interior and exterior regions referred as $V^-$ and $V^+$
respectively. The interior spacetime to $\Sigma$ can be described
by the line element
\begin{equation}\label{1}
ds_-^2=dt^2-X^2dr^2-Y^2(d\theta^2+\sin^2\theta d\phi^2),
\end{equation}
where $X$ and $Y$ are functions of $t$ and $r$. For the exterior
spacetime to $\Sigma$, we take the Schwarzschild spacetime
\begin{equation}\label{2}
ds_+^2=(1-\frac{2M}{R})dT^2-\frac{1}{1-\frac{2M}{R}}dR^2-R^2(d\theta^2+\sin^2\theta
d\phi^2),
\end{equation}
where $M$ is a constant. In accordance with Israel junction
conditions \cite{9}, we suppose that the first and second
fundamental forms inherited by $\Sigma$ from the interior and
exterior spacetimes are same, i.e.,
\begin{enumerate}
\item The continuity of line element over $\Sigma$ gives
\begin{equation}\label{4}
(ds^2_-)_{\Sigma}=(ds^2_+)_{\Sigma}=ds^2_{\Sigma}.
\end{equation}
\item The continuity of extrinsic curvature over $\Sigma$
yields
\begin{equation}\label{5}
[K_{ij}]=K^+_{ij}-K^-_{ij}=0,\quad(i,j=0,2,3)
\end{equation}
where $K_{ij}$ is the extrinsic curvature defined as
\end{enumerate}
\begin{equation}\label{6}
K^{\pm}_{ij}=-n^{\pm}_{\sigma}(\frac{{\partial}^2x^{\sigma}_{\pm}}
{{\partial}{\xi}^i{\partial}{\xi}^j}+{\Gamma}^{\sigma}_{{\mu}{\nu}}
\frac{{{\partial}x^{\mu}_{\pm}}{{\partial}x^{\nu}_{\pm}}}
{{\partial}{\xi}^i{\partial}{\xi}^j}),\quad({\sigma},
{\mu},{\nu}=0,1,2,3).
\end{equation}
Here $\xi^i$ correspond to the coordinates on ${\Sigma }$,
$x^{\sigma}_{\pm}$ are the coordinates in $V^\pm$, the Christoffel
symbols $\Gamma^{\sigma}_{{\mu}{\nu}}$ are calculated from the
interior or exterior spacetimes and $n^{\pm}_{\sigma}$ are
components of outward unit normals to ${\Sigma}$ in the
coordinates $x^{\sigma}_{\pm}$.

In terms of interior and exterior spacetime coordinates
respectively, the equation of hypersurface  is given as follows
\begin{eqnarray}\label{8}
h_-(r,t)=r-r_{\Sigma}=0,\\
\label{9} h_+(R,T)=R-R_{\Sigma}(T)=0,
\end{eqnarray}
where $r_{\Sigma}$ is a constant. When we make use of Eq.(\ref{8})
in Eq.(\ref{1}), the metric on $\Sigma$ becomes
\begin{equation}\label{10}
(ds_-^2)_\Sigma={dt^2-Y^2(r_\Sigma
,t)(d\theta^2+\sin\theta^2d\phi^2)}.
\end{equation}
Also, Eqs.(\ref{9}) and (\ref{2}) yield
\begin{equation}\label{11}
(ds_+^2)_\Sigma=\left(1-\frac{2M}{R_\Sigma}-\frac{1}{1-\frac{2M}{R_\Sigma}}
(\frac{dR_\Sigma}{dT})^2\right)dT^2-R_\Sigma^2(d\theta^2+\sin\theta^2d\phi^2).
\end{equation}
Here we assume that
\begin{equation}\label{12}
\left(1-\frac{2M}{R_\Sigma}-\frac{1}{1-\frac{2M}{R_\Sigma}}
(\frac{dR_\Sigma}{dT})^2\right)>0,
\end{equation}
so that T is a timelike coordinate. From Eqs.(\ref{4}), (\ref{10})
and (\ref{11}), it follows that
\begin{eqnarray}\label{13}
R_\Sigma=Y(r_\Sigma,t),\\\label{14}
\left(1-\frac{2M}{R_\Sigma}-\frac{1}{1-\frac{2M}{R_\Sigma}}
(\frac{dR_\Sigma}{dT})^2\right)^{\frac{1}{2}}dT=dt .
\end{eqnarray}

The outward unit normals in regions $V^-$ and $V^+$, respectively,
are found from Eqs.(\ref{8}) and (\ref{9})
\begin{eqnarray}\label{15}
n^-_\mu&=&(0,X(r_\Sigma,t),0,0),\\
\label{16} n^+_\mu&=&(-\dot{R}_\Sigma,\dot{T}, 0,0).
\end{eqnarray}
The components of extrinsic curvature $K^\pm_{ij}$ become
\begin{eqnarray}\label{17}
K^-_{00}&=&0,\\
\label{18}
K^-_{22}&=&\csc^2{\theta}K^-_{33}=\left(\frac{YY'}{X}\right)_\Sigma,\\
\label{19}
K^+_{00}&=&\left(\dot{R}\ddot{T}-\dot{T}\ddot{R}+\frac{3M\dot{R}^2\dot{T}}{R(R-2M)}
-\frac{M(R-2M)\dot{T}^3}{R^3}
\right)_\Sigma,\\
\label{20} K^+_{22}&=&\csc^2{\theta}
K^+_{33}=[\dot{T}(R-2M)]_{\Sigma},
\end{eqnarray}
where dot and prime mean differentiation with respect to $t$ and $r$
respectively. From Eq.(\ref{5}), the continuity of extrinsic
curvature gives
\begin{eqnarray}\label{21}
K^+_{00}=0,\quad K^+_{22}=K^-_{22}.
\end{eqnarray}
Using Eqs.(\ref{17})-(\ref{21}) along with Eqs.(\ref{13}) and
(\ref{14}), the junction conditions turn out to be
\begin{eqnarray}\label{23}
(X\dot{{Y'}}-\dot{X}{Y'})_\Sigma=0,\\
\label{24} M=\left(\frac{Y}{2}
+\frac{Y}{2}\dot{Y}^2-\frac{Y}{2X^2}{Y'}^2\right)_{\Sigma}.
\end{eqnarray}
Equations (\ref{13}), (\ref{14}), (\ref{23}) and (\ref{24}) are
the necessary and sufficient conditions for the smooth matching of
the interior and exterior regions.

\section{Field Equations in $f(R)$ Gravity and their Solution}

The action of $f(R)$ gravity in the presence of matter described
by the matter Lagrangian $L_{M}$ is given by \cite{1}
\begin{equation}\label{1.11.1}
S=\int d^{4}x\sqrt{-g}\left[\frac{f(R)}{2\kappa}+L_{M}\right],
\end{equation}
where the matter Lagrangian depends on the metric $g_{\mu\nu}$ and
the matter fields. Varying this action with respect to metric
tensor, we obtain the following field equations
\begin{equation}\label{1.11.2}
F(R) R_{\mu\nu} - \frac{1}{2}f(R)g_{\mu\nu}-\nabla_{\mu}
\nabla_{\nu}F(R)+ g_{\mu\nu} \Box F(R)= \kappa T_{\mu\nu},
\end{equation}
where $F(R)\equiv df(R)/dR,~\Box \equiv \nabla^{\mu}\nabla_{\mu}$
with $\nabla_{\mu}$ representing the covariant derivative and
$\kappa(=8\pi)$ is the coupling constant in gravitational units.
These are the fourth order partial differential equations in the
metric tensor. The fourth order is due to the last two terms on
the left side of the equation. Capozziello, \textit{et al}. have
discussed the general solutions of these fourth order equation in
the context of spherically symmetry \cite{capo}. If we take
$f(R)=R$, these equations reduce to the field equations in general
relativity. The energy-momentum tensor for perfect fluid is
\begin{equation}\label{26}
{T_{{\mu}{\nu}}={({\rho}+p)}u_{\mu}u_{\nu}-pg_{\mu\nu}},
\end{equation}
where $\rho$ is the energy density, $p$ is the pressure and
$u_\mu=\delta^0_\mu$ is the four-vector velocity in co-moving
coordinates. Using Eq.(\ref{26}) in Eq.(\ref{1.11.2}), we have
\begin{equation}\label{28}
R_{{\mu}{\nu}}=\frac{1}{F(R)}[8\pi(({\rho}+p)u_{\mu}u_{\nu}-pg_{{\mu}{\nu}})
+ \frac{1}{2}f(R)g_{\mu\nu}+\nabla_{\mu} \nabla_{\nu}F(R)-
g_{\mu\nu} \Box F(R)] .
\end{equation}
Equations (\ref{28}) for the interior spacetime take the form
\begin{eqnarray} \label{f1}
R_{00}&=&-\frac{\ddot{X}}{X}-2\frac{\ddot{Y}}{Y}=\frac{1}{F(R)}[8\pi\rho
+
\frac{1}{2}f(R)-(\frac{-F''(R)}{X^2}+\frac{\dot{X}\dot{F}(R)}{X}\nonumber\\
&+&\frac{\dot{X}F'(R)}{X^3}
+\frac{2\dot{Y}\dot{F}(R)}{Y}-\frac{2Y'F'(R)}{X^2Y})],\\
\label{f2}
R_{11}&=&-\frac{\ddot{X}}{X}-2\frac{\dot{X}}{X}\frac{\dot{Y}}{Y}
+\frac{2}{X^2}[\frac{Y''}{Y}-\frac{Y'X'}{XY}]\nonumber\\
&=&\frac{1}{F(R)}[\frac{1}{2}f(R)- 8\pi p+X^2(\ddot{F}(R)
+\frac{2\dot{Y}\dot{F}(R)}{Y}-\frac{2Y'F'(R)}{X^2Y})] ,\\
\label{f3}
R_{22}&=&-\frac{\ddot{Y}}{Y}-(\frac{\dot{Y}}{Y})^2
-\frac{\dot{X}}{X}\frac{\dot{Y}}{Y}
+\frac{2}{X^2}[\frac{Y''}{Y}+(\frac{Y'}{Y})^2-
\frac{X'}{X}\frac{Y'}{Y}-(\frac{X}{Y})^2]\nonumber\\
&=&\frac{1}{F(R)}[\frac{1}{2}f(R)-8\pi
p+Y^2(\ddot{F}(R)-\frac{F''(R)}{X^2}+\frac{\dot{X}\dot{F}(R)}{X}
+\frac{\dot{X}F'(R)}{X^3})],\nonumber\\
\end{eqnarray}
\begin{eqnarray}
\label{f4}
R_{33}&=&{\sin}^2{\theta}R_{22},\\
\label{f5}
R_{01}&=&-2\frac{\dot{Y'}}{Y}+2\frac{\dot{X}}{X}\frac{Y'}{Y}=\frac{1}{F(R)}[\dot{F'}-\frac{\dot
X}{X} F'] .
\end{eqnarray}
To solve this set of the field equations (\ref{f1})-(\ref{f5}), we
need to integrate Eq.(\ref{f5}) to get the explicit value of $X$. It
follows from Eq.(\ref{f5}) that
\begin{equation*}
X=\int\frac{2\dot{Y'}F+\dot{F'}Y}{2Y'F+F'Y}dt .
\end{equation*}
We see that this cannot be integrated unless we have some value of
$F$ such that derivative of the denominator should be the numerator.
The condition of constant scalar curvature ($R=R_{0}$), according to
which $F(R_0)=constant$, enables us to solve this integral. This
means that the Ricci scalar obtained by contracting Eq.(\ref{28})
must be constant which is possible only if we take pressure and
density constant, i.e., $p=p_0,~\rho=\rho_0$. Consequently, the
field equations become
\begin{eqnarray} \label{42}
&&-\frac{\ddot{X}}{X}-2\frac{\ddot{Y}}{Y}=\frac{1}{F(R_0)}[8\pi\rho_0
+ \frac{1}{2}f(R_0)],\\
\label{43}
&&-\frac{\ddot{X}}{X}-2\frac{\dot{X}}{X}\frac{\dot{Y}}{Y}
+\frac{2}{X^2}[\frac{Y''}{Y}-\frac{Y'X'}{XY}]
=\frac{1}{F(R_0)}[\frac{1}{2}f(R_0)- 8\pi p_0] ,\\
\label{44} &&-\frac{\ddot{Y}}{Y}-(\frac{\dot{Y}}{Y})^2
-\frac{\dot{X}}{X}\frac{\dot{Y}}{Y}
+\frac{2}{X^2}[\frac{Y''}{Y}+(\frac{Y'}{Y})^2-
\frac{X'}{X}\frac{Y'}{Y}-(\frac{X}{Y})^2]\nonumber\\
&&=\frac{1}{F(R_0)}[\frac{1}{2}f(R_0)- 8\pi p_0] ,\\
\label{46} &&\frac{\dot{Y'}}{Y}-\frac{\dot{X}}{X}\frac{Y'}{Y}=0 .
\end{eqnarray}

Now we solve Eqs.(\ref{42})-(\ref{46}). Integration of Eq.(\ref{46})
with respect to $t$ yields
\begin{equation}\label{47}
X=\frac{Y'}{W},
\end{equation}
where $W=W(r)$ is an arbitrary function of $r$. Using this
equation in Eqs.(\ref{42})-(\ref{46}), it follows that
\begin{equation}\label{50}
2\frac{\ddot{Y}}{Y}+(\frac{\dot{Y}}{Y})^2+\frac{(1-W^2)}{Y^2}
=\frac{1}{F(R_0)}\left[4 \pi
(p_0-\rho_0)-\frac{1}{2}f(R_0)\right].
\end{equation}
Integrating this equation with respect to $t$, it follows that
\begin{equation}\label{54}
{\dot{Y}}^2=W^2-1+2\frac{m}{Y}+\frac{1}{F(R_0)}\left[4 {\pi}
(p_0-\rho_0)-\frac{1}{2}f(R_0)\right]\frac{Y^2}{3},
\end{equation}
where $m=m(r)$ is an arbitrary function of $r$ and is related to the
mass of the collapsing system. Substituting Eqs.(\ref{47}),
(\ref{54}) into Eq.(\ref{42}), we get
\begin{equation}\label{55}
m'=\frac{4\pi}{F(R_0)}(p_0+\rho_0){{Y'}{Y^2}}.
\end{equation}
Integrating this equation with respect to $r$, we obtain
\begin{equation}\label{56}
m(r)=\frac{4\pi}{F(R_0)}(p_0+\rho_0)\int{Y'Y^2}dr + c(t),
\end{equation}
where $c(t)$ is a function of integration. Since mass cannot be
negative due to physical reasons, we must take the function $m(r)$
positive. Using Eqs.(\ref{47}) and (\ref{54}) into the junction
condition Eq.(\ref{24}), it follows that
\begin{equation}\label{57}
M=m+\frac{1}{6F(R_0)}\left[4\pi
(p_0-\rho_0)-\frac{1}{2}f(R_0)\right]Y^3.
\end{equation}
The total energy $\tilde{M}(r,t)$ up to a radius $r$ at time $t$
inside the hypersurface $\Sigma$ can be calculated by using the
definition of mass function \cite{16} given by
\begin{equation}\label{58}
\tilde{M}(r,t)=\frac{1}{2}Y(1+g^{\mu\nu}Y_{,\mu} Y_{,\nu}).
\end{equation}
For the interior metric, it becomes
\begin{equation}\label{59}
\tilde{M}(r,t)=\frac{1}{2}Y[1+\dot{Y}^2-(\frac{Y'}{X})^2].
\end{equation}
Replacing Eqs.(\ref{47}) and (\ref{54}) in Eq.(\ref{59}), we obtain
\begin{equation}\label{60}
\tilde{M}(r,t)=m(r)+\frac{1}{6F(R_0)}\left[4\pi
(p_0-\rho_0)-\frac{1}{2}f(R_0)\right]Y^3.
\end{equation}
Here we assume
\begin{equation}\label{61*}
\frac{1}{F(R_0)}\left[4\pi (p_0-\rho_0)-\frac{1}{2}f(R_0)\right]>0
\end{equation}
and the condition
\begin{equation}\label{61}
W(r)=1.
\end{equation}
Using Eqs.(\ref{47}), (\ref{54}) and (\ref{61}) so that we have
analytic solutions in closed form as follows
\begin{eqnarray}\label{62}
Y&=&\left(\frac{6m F(R_0)}{4\pi
(p_0-\rho_0)-\frac{1}{2}f(R_0)}\right)^\frac{1}{3}\sinh^\frac{2}{3}\alpha(r,t),\\
\label{63} X&=&\left(\frac{6m F(R_0)}{4\pi
(p_0-\rho_0)-\frac{1}{2}f(R_0)}\right)^\frac{1}{3}[\frac{m'}{3m}\sinh\alpha(r,t)\nonumber\\
&+&t'_s(r)\sqrt{\frac{4\pi
(p_0-\rho_0)-\frac{1}{2}f(R_0)}{3F(R_0)}}\cosh\alpha(r,t)]
\sinh^\frac{-1}{3}\alpha(r,t),
\end{eqnarray}
where
\begin{equation}\label{64}
\alpha(r,t)=\sqrt{\frac{3(4\pi
(p_0-\rho_0)-\frac{1}{2}f(R_0))}{4F(R_0)}}[t_s(r)-t].
\end{equation}
Here $t_s(r)$ is an arbitrary function of $r$ which denotes the time
formation of singularity for a particular shell at distance $r$.

When $f(R_0)\rightarrow8\pi (p_0-\rho_0)$, the above solution
corresponds to the Tolman-Bondi solution \cite{17}
\begin{eqnarray}\label{65}
\lim_{f(R_0)\rightarrow8\pi
(p_0-\rho_0)}X(r,t)&=&\frac{m'(t_s-t)+2mt'_s}
{[6m^2(t_s-t)]^\frac{1}{3}},\\
\label{66} \lim_{f(R_0)\rightarrow8\pi (p_0-\rho_0)}Y(r,t)&=&
\left[\frac{9m}{2}(t_s-t)^2\right]^{\frac{1}{3}}.
\end{eqnarray}

\section{Apparent Horizons}

The apparent horizon is formed when the boundary of trapped two
spheres is formed. The formula to find such a boundary with null
outward normals is given as follows
\begin{equation}\label{67}
g^{\mu\nu}Y_{,\mu} Y_{,\nu}=\dot{Y}^2-(\frac{Y'}{X})^2=0.
\end{equation}
Using Eqs.(\ref{47}) and (\ref{54}), we get
\begin{equation}\label{68}
\frac{1}{F(R_0)}\left[4\pi
(p_0-\rho_0)-\frac{1}{2}f(R_0)\right]Y^3-3Y+6m=0.
\end{equation}
The values of $Y$ yield the apparent horizons. For $f(R_0)=8\pi
(p_0-\rho_0)$, we have $Y=2m$, i.e., Schwarzschild horizon.
Equation (\ref{68}) can have the
following positive roots.\\\\
\textbf{Case (i)}: For $3m<\sqrt{\frac{F(R_0)}{4\pi
(p_0-\rho_0)-\frac{1}{2}f(R_0)}}$, we obtain two horizons, namely
cosmological horizon, $Y_c$, and black hole horizon, $Y_{bh}$,
i.e.
\begin{eqnarray}\label{69}
Y_c&=&\sqrt{\frac{4F(R_0)}{4\pi
(p_0-\rho_0)-\frac{1}{2}f(R_0)}}\cos\frac{\varphi}{3},\\
\label{70} Y_{bh}&=&-\sqrt{\frac{4F(R_0)}{4\pi
(p_0-\rho_0)-\frac{1}{2}f(R_0)}}
\left(\cos\frac{\varphi}{3}-\sqrt{3}\sin\frac{\varphi}{3}\right),
\end{eqnarray}
where
\begin{equation}\label{71}
\cos\varphi=-3m\sqrt{\frac{F(R_0)}{4\pi
(p_0-\rho_0)-\frac{1}{2}f(R_0)}}.
\end{equation}
If we take $m=0$, it follows from Eqs.(\ref{69}) and (\ref{70})
that
\begin{equation}\label{71*}
Y_c=\sqrt{\frac{3F(R_0)}{4\pi
(p_0-\rho_0)-\frac{1}{2}f(R_0)}}~~\textmd{and}~~ Y_{bh}=0.
\end{equation}
For $m\neq0$ and $f(R_0)\neq 8\pi (p_0-\rho_0)$, $Y_c$
and $Y_{bh}$ can be generalized \cite{18} respectively.\\\\
\textbf{Case (ii):} For $3m=\sqrt{\frac{F(R_0)}{4\pi
(p_0-\rho_0)-\frac{1}{2}f(R_0)}}$, there is only one positive root
which corresponds to a single horizon i.e.,
\begin{equation}\label{72}
Y_c=Y_{bh}=\sqrt{\frac{F(R_0)}{4\pi
(p_0-\rho_0)-\frac{1}{2}f(R_0)}}.
\end{equation}
This shows that both horizons coincide. The ranges for these
horizons can be written as follows
\begin{equation}\label{73}
0\leq Y_{bh} \leq \sqrt{\frac{F(R_0)}{4\pi
(p_0-\rho_0)-\frac{1}{2}f(R_0)}} \leq Y_{c} \leq
\sqrt{\frac{3F(R_0)}{4\pi (p_0-\rho_0)-\frac{1}{2}f(R_0)}}.
\end{equation}
The largest proper area of the black hole horizon is given by
\begin{equation}\label{72*}
{4\pi}Y^2=\frac{8 \pi F(R_0)}{8\pi p_{0}(p_0-\rho_0)-f(R_0)}
\end{equation}
and the cosmological horizon has its area between
\begin{equation}\label{72**}
\frac{8 \pi F(R_0)}{8\pi (p_0-\rho_0)-f(R_0)} ~~\textmd{and}~~
\frac{24 \pi F(R_0)}{8\pi (p_0-\rho_0)-f(R_0)}\
\end{equation}
\textbf{Case (iii):} For $3m>\sqrt{\frac{F(R_0)}{4\pi
(p_0-\rho_0)-\frac{1}{2}f(R_0)}}$, there are no positive roots and
consequently there are no apparent horizons.

Now we calculate the time of formation for the apparent horizon
using Eqs.(\ref{62}) and (\ref{68})
\begin{equation}\label{74}
t_n=t_s-\sqrt{\frac{4 F(R_0)}{12\pi
(p_0-\rho_0)-\frac{3}{2}f(R_0)}}\sinh^{-1}
(\frac{Y_n}{2m}-1)^{\frac{1}{2}}, \quad(n=1,2).
\end{equation}
When $f(R_0)\rightarrow8\pi (p_0-\rho_0)$, this corresponds to
Tolman-Bondi solution \cite{17} given as
\begin{equation}\label{75}
t_{ah}=t_s-\frac{4}{3}m.
\end{equation}
From Eq.(\ref{74}), we can write
\begin{equation}\label{76}
\frac{Y_n}{2m}=\cosh^{2}\alpha_n,
\end{equation}
where
\begin{equation}\label{76*}
\alpha_n(r,t)=\sqrt{\frac{3(4\pi
(p_0-\rho_0)-\frac{1}{2}f(R_0))}{4F(R_0)}}[t_s(r)-tn)].
\end{equation}
Equations (\ref{73}) and (\ref{74}) imply that $Y_{c}\geq Y_{bh}$
and $t_{2} \geq t_{1}$ respectively. Here $t_1$ denotes the time
formation of cosmological horizon and $t_2$ denotes the time
formation of black hole horizon. The inequality $t_{2} \geq t_{1}$
indicates that the cosmological horizon forms earlier than the
black hole horizon.

The time difference between the formation of cosmological horizon
and singularity and the formation of black hole horizon and
singularity respectively can be found as follows. Using
Eqs.(\ref{69})-(\ref{71}), it follows that
\begin{eqnarray}\label{77}
\frac{d(\frac{Y_c}{2m})}{dm}&=&\frac{1}{m}\left(-\frac{\sin\frac{\varphi}{3}}{\sin\varphi}
+\frac{3\cos\frac{\varphi}{3}}{\cos\varphi}\right)<0,\\
\label{78}\frac{d(\frac{Y_{bh}}{2m})}{dm}
&=&\frac{1}{m}\left(-\frac{\sin\frac{(\varphi+4\pi)}{3}}{\sin\varphi}
+\frac{3\cos\frac{(\varphi+4\pi)}{3}}{\cos\varphi}\right)>0.
\end{eqnarray}
The time difference between the formation of singularity and
apparent horizons is
\begin{equation}\label{79}
T_n=t_s-t_n.
\end{equation}
It follows from Eq.(\ref{76}) that
\begin{equation}\label{3.1.18}
\frac{dT_n}{d(\frac{Y_n}{2m})}
=\frac{1}{\sinh\alpha_n\cosh\alpha_n{\sqrt{\frac{3}{F(R_0)}[4\pi
(p_0-\rho_0) -\frac{1}{2}f(R_0)]}}}.
\end{equation}
Using Eqs.(\ref{77}) and (\ref{3.1.18}), we get
\begin{eqnarray}\label{81}
\frac{dT_1}{d
m}=\frac{dT_1}{d(\frac{Y_c}{2m})}\frac{d(\frac{Y_c}{2m})}{d m}
=\frac{1}{m{\sqrt{\frac{3}{F(R_0)}[4\pi
(p_0-\rho_0) -\frac{1}{2}f(R_0)]}}\sinh\alpha_1\cosh\alpha_1}\nonumber\\
\times\left(-\frac{\sin\frac{\varphi}{3}}{\sin\varphi}
+\frac{3\cos\frac{\varphi}{3}}{\cos\varphi}\right)<0 .
\end{eqnarray}
It shows that $T_1$ is a decreasing function of mass $m$. This means
that time interval between the formation of cosmological horizon and
singularity is decreased with the increase of mass. Similarly, from
Eqs.(\ref{78}) and (\ref{3.1.18}), we get
\begin{eqnarray}\label{82}
\frac{dT_2}{dm}= \frac{1}{m{\sqrt{\frac{3}{F(R_0)}[4\pi
(p_0-\rho_0) -\frac{1}{2}f(R_0)]}}\sinh\alpha_2\cosh\alpha_2}\nonumber\\
\times\left(-\frac{\sin\frac{(\varphi+4\pi)}{3}}{\sin\varphi}
+\frac{3\cos\frac{(\varphi+4\pi)}{3}}{\cos\varphi}\right)>0.
\end{eqnarray}
This indicates that $T_2$ is an increasing function of mass $m$
showing that the time difference between the formation of black hole
horizon and singularity is increased with the increase of mass.

\section{Summary and Discussion}

In this paper we have discussed different aspects of gravitational
collapse in metric $f(R)$ gravity. For this purpose, the field
equations are solved using the assumption of constant scalar
curvature. To meet the requirement of gravitational collapse, we
assume that scalar curvature is very large constant quantity. We
obtain two physical apparent horizons (Eq.(\ref{68})) named as black
hole horizon and cosmological horizon. It is concluded that black
hole horizon requires more time for its formation as compared to
cosmological horizon. However, both the horizons form earlier than
singularity which indicates that singularity is covered, i.e, black
hole. In this way, $f(R)$ gravity seems to support CCC.

Now we discuss how the constant scalar curvature term $f(R_0)$
affects the rate of gravitational collapse. For the exterior metric,
the Newtonian potential, $\phi=\frac{1}{2}(1-g_{00})$, becomes
\begin{equation}\label{84}
\phi(R)=\frac{m}{R}+\frac{R^2}{6F(R_0)}\left[4\pi (p_0-\rho_0)
-\frac{1}{2}f(R_0)\right].
\end{equation}
Taking derivative of this equation with respect to $R$, we obtain
the corresponding Newtonian force
\begin{equation}\label{85}
F=-\frac{m}{R^2}+\frac{R}{3F(R_0)}\left[4\pi (p_0-\rho_0)
-\frac{1}{2}f(R_0)\right].
\end{equation}
This force has no effect on collapsing matter for the following
values of $m$ and $R$
\begin{eqnarray}\nonumber
m=\frac{1}{3\sqrt{\frac{1}{F(R_0)}[4\pi
(p_0-\rho_0)-\frac{1}{2}f(R_0)]}}\\\nonumber
R=\frac{1}{\sqrt{\frac{1}{F(R_0)}[4\pi (p_0-\rho_0)
-\frac{1}{2}f(R_0)]}}.
\end{eqnarray}
However, the force will be repulsive (positive) if the values of
$m$ and $R$ are larger than the above values, provided that
$\frac{1}{F(R_0)}[4\pi(p_0-\rho_0) -\frac{1}{2}f(R_0)]>0$. In
other words, the repulsive force can be generated by $f(R_0)$ if
$f(R_0)< 8\pi (p_0-\rho_0)$ over the entire range of the
collapsing sphere. Hence, this repulsive force slows down the
collapse of matter. The case of attractive (negative) force
contradicts Eq.(\ref{61*}) and hence our results do not favor the
faster rate of the collapsing process. The above results can also
be verified from Eq.(\ref{54}) indicating the acceleration of
collapsing matter as follows
\begin{equation}\label{86}
\ddot{Y}=-\frac{m}{Y^2}+\frac{1}{F(R_0)}\left[4\pi (p_0-\rho_0)
-\frac{1}{2}f(R_0)\right]\frac{Y}{3}.
\end{equation}
The analysis of positive and negative acceleration provides the
same results on $f(R_0)$ as we obtained from the Newtonian force.

We would like to mention here why to explore the constant
curvature solution in a typical modified modified $f(R)$ gravity.
To generate the accelerating expansion in the present universe,
$f(R)$ could be a small constant \cite{31}. The universe starts
from the inflation driven by the effective cosmological constant
at the early stage, where curvature is very large. When the
curvature becomes smaller, the effective cosmological constant
also becomes smaller. There appears the small effective
cosmological constant when the density of the radiation and the
matter becomes small and the curvature goes to the value $R_0$.
Thus expansion could start and cosmological constant can be
identified as $f(R_0)$ in the present accelerating era.

It is well-known that only those $f(R)$ models are cosmologically
viable whose solution correspond to general relativity. The class of
constant scalar curvature solutions with spherically symmetry
contains black hole solutions in the presence of cosmological
constant. For example, the Schwarzschild anti de-Sitter solutions
and all the topological solutions associated with a negative
$\Lambda_{eff}$. Exact solutions of cylindrical symmetric spacetimes
with assumptions of constant scalar curvature are applicable to the
exterior of a string.

Finally, we conclude that the term $f(R_0)$ plays the same role as
that of the cosmological constant in the Einstein field equations
\cite{12}. In general relativity, the cosmological constant acts
as a repulsive force (hence slows down the collapse of matter) and
same does $f(R_0)$. It would be worthwhile to investigate these
issues in $f(R)$ theory by dropping the assumption of scalar
curvature for complete understanding of gravitational collapse.

\vspace{0.25cm}

{\bf Acknowledgment}

\vspace{0.25cm}

We would like to thank the Higher Education Commission, Islamabad,
Pakistan for its financial support through the {\it Indigenous
Ph.D. 5000 Fellowship Program Batch-III}.


\begin{thebibliography}{40}
\bibitem{1} Sahani, V. and Starobinsky, A.A.: Int. J. Mod. Phys. \textbf{D9}(2000)373.

\bibitem{2} Sahani, V. and Starobinsky, A.A.: Int. J. Mod.
Phys. \textbf{D15}(2006)2195;\\
Sami, M.: Lecture Notes Phys. \textbf{72}(2007)219;\\
Peebles, P.J.E. and Ratra, B.: Rev. Mod. Phys. \textbf{75}(2003)559.

\bibitem{3} Copeland, E.J., Sami, M. and Tsujikawa, S.: Int. J. Mod. Phys. \textbf{D15}(2006)1753.

\bibitem{1*} Capozziello, S.: Int. J. Mod. Phys.
\textbf{D11}(2002)483;\\
Abdalla, M.C.B., Nojiri, S. and Odintsov,
S.D.: Class. Quantum Grav. \textbf{22}(2005)L35;\\
Sotiriou, T.P.:
Class. Quantum Grav. \textbf{23}(2006)5117.

\bibitem{2*} Nojiri, S. and Odintsov,
S.D.: Phys. Lett. \textbf{B646}(2007)105; \emph{ibid}. \textbf{B659}(2008)821;\\
Bertolami, O. and P\'{a}rmos, J.: Class. Quantum Grav. \textbf{25}(2008)245017;\\
Santose, J. Alcaniz, J.S., Carvalho, F.C. and Pires, N.: Phys.
Lett. \textbf{B669}(2008)14.

\bibitem{3*} Capozziello, S. and Francaviglia, M.: Gen. Rel. Gravit.
\textbf{40}(2008)357;\\
Harko, T.: Phys. Lett. \textbf{B669}(2008)376;\\ Dev, A.
\textsl{et al.}: Phys. Rev. \textbf{D78}(2008)083515.

\bibitem{4*} Chiba, T.: Phys. Lett. \textbf{B575}(2003)1;\\
Amendola, L. and Tsujikawa, S.: Phys. Lett.
\textbf{B660}(2008)125;\\
Lecian, O.M. and  Montani, G.: Class. Quantum Grav.
\textbf{26}(2009)045014.

\bibitem{5*} Dicke, R.: Gen. Rel. Gravit.
\textbf{36}(2004)217;\\ Sotiriou, T.P.: Phys. Rev.
\textbf{D73}(2006)063515; Gen. Rel. Gravit. \textbf{38}(2006)1407.

\bibitem{6*} Dolgov, A.D. and Kawasaki, M.: Phys. Lett. \textbf{B573}(2003)1;\\
Faraoni, V.: Phys. Rev. \textbf{D74}(2006)104017;\\
Sotiriou, T.P.: Phys. Lett. \textbf{B645}(2007)389.

\bibitem{7*} Frolov, A.V.: Phys. Rev. Lett. \textbf{101}(2008)061103;

\bibitem{7**} Kung, J.H.: Phys. Rev. \textbf{D53}(1996)3017;\\
Bergliaffa, S.E.P.: Phys. Lett. \textbf{B642}(2006)311;\\
Bertolami, O. and Sequeira, M.C.: Phys. Rev.
\textbf{D79}(2009)104010.

\bibitem{6}Amendola, L., Polarski, D. and Tsujikawa S.: Phys. Rev. Lett.
\textbf{98}(2007)131302.

\bibitem{7} Erickcek, A.L., Smith, T.L. and Kamionkowski, M.:
Phys. Rev. \textbf{D74}(2006)121501.

\bibitem{8} Kainulainen, K., Reijonen, V. and Sunhede, D.: Phys. Rev. \textbf{D76}(2007)043503.

\bibitem{os} Oppenheimer, J.R. and Synder, H.: Phys. Rev.
\textbf{56}(1939)455.

\bibitem{Herr} Herrera, L. and Santos, N.O.: Phys. Rev. {\bf
D70}(2004)084004.

\bibitem{10**} Hawking, S.W. and Ellis, G.F.R.: \emph{The Large Scale Structure of Spacetime}
(Cambridge University Press, Cambridge, 1975).

\bibitem{11*} Penrose, R.: Phys. Rev. Lett. \textbf{14}(1965)57;\\
Hawking, S.W.: Proc. R. Soc. London \textbf{A300}(1967)187;\\
Hawking, S.W. and Penrose, R.: Proc. R. Soc. London
\textbf{A314}(1970)529.

\bibitem{11} Penrose, R.: Riv. Nuovo Cimento \textbf{1}(1969)252.

\bibitem{CC} Szekers, P.: Commun. Math. Phys.
\textbf{41}(1975)55;\\ Joshi, P.S. and Dwivedi, I.H.: Commun.
Math. Phys. \textbf{166}(1994)117; Class. Quantum Grav.
\textbf{16}(1999)41;\\ Lake, K.: Phys. Rev. Lett.
\textbf{68}(1992)3129;\\ Ori, A. and Piran, T.: Phys. Rev. Lett.
\textbf{59}(1987)2137;\\ Harada, T.: Phys. Rev.
\textbf{D58}(1998)104015.

\bibitem{CCC} Joshi, P.S.: \emph{Global Aspects in Gravitation and Cosmology}
(Oxford University Press, Oxford, 1993).

\bibitem{MS} Markovic, D. and Shapiro, S.L.: Phys. Rev.
\textbf{D61}(2000)084029.

\bibitem{Lake} Lake, K.: Phys. Rev. \textbf{D62}(2000)027301.

\bibitem{12} Sharif, M. and Ahmad, Z.: Mod. Phys. Lett.
\textbf{A22}(2007)1493; \emph{ibid.} \textbf{A22}(2007)2947.

\bibitem{hd} Sharif, M. and Ahmad, Z.: J. Korean Physical Society 52(2008)980;
Acta Physica Polonica \textbf{B39}(2008)1337.

\bibitem{sa} Sharif, M. and Abbas, G.: Mod.
Phys. Lett. \textbf{A}(2009, to appear).

\bibitem{capo} Capozziello, S., Stabile, A., Troisi, A.: Class. Quantum
Grav. \textbf{25}(2008)085005.

\bibitem{9} Israel, W.: Nuovo Cimento \textbf{B44}(1966)1; \emph{ibid.}
\textbf{B48}(1967)463(E).

\bibitem{16} Misner, C.W. and Sharp, D.: Phys. Rev. \textbf{136}(1964)b571.

\bibitem{17} Eardley, D.M. and Smarr, L.: Phys. Rev.
\textbf{D19}(1979)2239.

\bibitem{18} Hayward, S.A., Shiromizu, T. and Nakao, K.: Phys. Rev.
\textbf{D49}(1994)5080.

\bibitem{31} Nojiri, S., Odintsov, S.D.: Phys. Lett.
\textbf{B657}(2007)238.
\end{thebibliography}
\end{document}